\documentclass[epj]{svjour}
\usepackage{latexsym}
\usepackage{graphics}
\begin{document}
\title{ Phase structure of a surface model with many fine holes}

\author{Hiroshi Koibuchi
}                     
%
%
\institute{Department of Mechanical and Systems Engineering \\
  Ibaraki National College of Technology \\
  Nakane 866, Hitachinaka, Ibaraki 312-8508, Japan }
%
%
\abstract{
We study the phase structure of a surface model by using the canonical Monte Carlo simulation technique on triangulated, fixed connectivity, and spherical surfaces with many fine holes. The size of a hole is assumed to be of the order of lattice spacing (or bond length) and hence can be negligible compared to the surface size in the thermodynamic limit. We observe in the numerical data that the model undergoes a first-order collapsing transition between the smooth phase and the collapsed phase. Moreover the Hasudorff dimension $H$ remains in the physical bound, i.e., $H<3$ not only in the smooth phase but also in the collapsed phase at the transition point. The second observation is that the collapsing transition is accompanied by a continuous transition of surface fluctuations. This second result distinguishes the model in this paper and the previous one with many holes, whose size is of the order of the surface size, because the previous surface model with large-sized holes has only the collapsing transition and no transition of surface fluctuations. 
}
\authorrunning {H.Koibuchi}
\titlerunning {Phase structure of a surface model with many fine holes}
\maketitle
\section{Introduction}\label{intro}
Biological membranes such as the so-called cell membranes or plasma membranes have many holes called transport protein or protein channel, through which some biological materials are transported in/out. Currently it is well known that cell membranes are heterogeneous due to cytoskeletons and membrane proteins including the transport proteins, and moreover, the membranes have complex structures such as rafts and fences \cite{Kusumi-JCB-1994,Kusumi-COCB-1996}. 

The fluid mosaic model \cite{Singer-Nicolson-1972} seems to be valid only when these structures of size smaller than the membrane size were neglected. By neglecting further the structures of size negligible compared to the membrane size, we have a homogeneous surface, which has been described by the conventional surface model of Helfrich and Polyakov \cite{HELFRICH-1973,POLYAKOV-NPB1986,KLEINERT-PLB1986}. The crumpling phenomena was extensively studied theoretically and numerically \cite{Peliti-Leibler-PRL1985,DavidGuitter-EPL1988,PKN-PRL1988,KANTOR-NELSON-PRA1987,Baum-Ho-PRA1990,CATTERALL-NPBSUP1991,AMBJORN-NPB1993,NISHIYAMA-PRE-2004} by using this homogeneous surface model. Current understanding of membranes on the basis of statistical mechanics are reviewed in \cite{NELSON-SMMS2004,Gompper-Schick-PTC-1994,Bowick-PREP2001,SEIFERT-LECTURE2004}. We also know that the collapsing transition is accompanied by a transition of surface fluctuations, and both transitions are of first-order \cite{KD-PRE2002,KOIB-PRE-2005,KOIB-NPB-2006}.

It was recently reported that a surface model with many holes undergoes a collapsing transition between the smooth phase and the collapsed phase, and no transition of surface fluctuations can be seen in that model \cite{KOIB-PRE-2007-1}. The results in \cite{KOIB-PRE-2007-1} are considered to be reflecting an inhomogeneous structure in biological membranes. The Hamiltonian of the model in \cite{KOIB-PRE-2007-1} is the one of Helfrich and Polyakov. 

However, the size of holes in \cite{KOIB-PRE-2007-1} is assumed to be of the order of the surface size in the limit of $N\to \infty$, ($N$ is the total number of vertices) or in other words in the thermodynamic limit. Thus, the phase structure of the surface model with small sized holes still remains to be studied. We expect that the phase structure of the homogeneous model is influenced by such holes, because the vertices at the edge of the holes are relatively freely moving; no bending energy is defined on the edges, i.e. those vertices are considered to be in a free boundary condition. For this reason, no transition of surface fluctuations can be seen in the model in \cite{KOIB-PRE-2007-1}. Thus, the influence of holes on the phase structure is expected to persist in such a case that the size of holes reduces sufficiently small compared to the surface size.  

In this paper, we study a surface model of Helfrich and Polyakov on triangulated, fixed connectivity and spherical surfaces with many fine holes by Monte Carlo (MC) simulations. The purpose of the present paper is to see whether or not the phase transitions of the conventional homogeneous model are influenced by the presence of many fine holes and change their properties. 

\section{Model}\label{model}
The partition function we are studying is of the form
\begin{eqnarray} 
\label{Part-Func}
 Z = \int^\prime \prod _{i=1}^{N} d X_i \exp\left[-S(X, {\cal T})\right],\\  
 S(X, {\cal T})=S_1 + b S_2. \nonumber
\end{eqnarray} 
The Hamiltonian $S$ is defined on triangulated lattices. The construction technique for the lattice with holes will be presented below. The symbols $X$ and ${\cal T}$ in $S$ denote the vertex position and the triangulation, where ${\cal T}$ is fixed in the simulations. $\int^\prime $ denotes that the center of mass of the surface is fixed. The Gaussian bond potential $S_1$ and the bending energy $S_2$ are defined so that
\begin{equation}
\label{Disc-Eneg} 
  S_1=\sum_{(ij)} (X_i-X_j)^2,  \quad S_2=\sum_{(ij)} (1-{\bf n}_i \cdot {\bf n}_j),
\end{equation} 
where the variable $X_i (\in {\bf R}^3)$ denotes the position of the vertex $i$, and ${\bf n}_i(\in {\bf S}^2$= the unit sphere in ${\bf R}^3$) denotes a unit normal vector of the triangle $i$. 

The construction technique for the triangulated lattice is almost identical to that for the lattices in \cite{KOIB-PRE-2007-1} and is summarized as follows: We start with the icosahedron, which is characterized by $12$ vertices, $30(\!=3\!\times\! 12\!-\!6)$ bonds, and $20(\!=\!2\!\times\! 12\!-\!4)$ triangles. By dividing the bonds of the icosahedron into $\ell$ pieces of uniform length $a$, we firstly have a triangulated spherical surface of size $N_0\!=\!10\ell^2\!+\!2$, which is the total number of vertices on the surface without holes. Secondly, a sublattice is obtained by dividing $\ell$ edges into $m$ partitions ($m\!=\!1,2,\cdots$), where each partition is of length $L\!=\!\ell /m$ in the unit of $a$ if $m$ divides $\ell$. The sublattice is identical to the compartment in \cite{KOIB-PRE-2007-2,KOIB-EPJB-2007-1}. Finally, we label one part of the compartments as holes and the remaining other part as the lattice points. Thus, we have a triangulated lattice with many holes. 

\begin{figure}[htb]
\resizebox{0.49\textwidth}{!}{%
\centering
\includegraphics{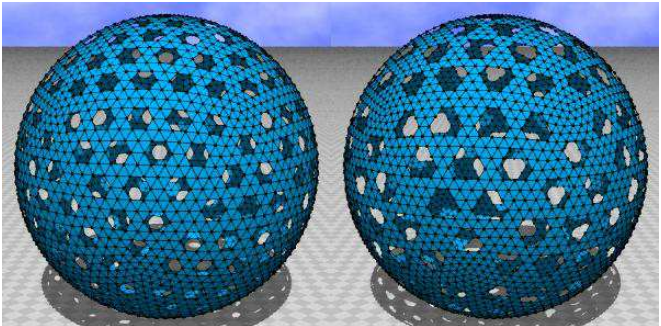}
}
\vspace{0.2cm}
\caption{ Surfaces with holes of size (a) $(N,L)\!=\!(2942,3)$ and (b) $(N,L)\!=\!(3402,4)$, which can also be characterized by two integers $(\ell,m)\!=\!(18,6)$ and $(\ell,m)\!=\!(20,5)$, respectively.
The holes in (a), (b) respectively contain $6$, $13$ triangles, which are excluded from the surface.
 } 
\label{fig-1}
\end{figure}
We should note that the size of holes can be characterized by $L$. Two types of lattices corresponding to
\begin{equation}
\label{lattices} 
  L=3, \quad L=4,
\end{equation} 
 are assumed in this paper. The minimum size of holes is given by the condition $L\!=\!3$, where the center of hole is a vertex that is excluded from the surface. The second minimum size is given by  $L\!=\!4$, where each hole has $3$ vertices that are excluded from the surface. Figures \ref{fig-1}(a) and \ref{fig-1}(b) show the lattices of $(N,L)\!=\!(2942,3)$ and $(N,L)\!=\!(3402,4)$, which are given by $(\ell,m)\!=\!(18,6)$ and $(\ell,m)\!=\!(20,5)$, respectively. The size of holes given by Eq.(\ref{lattices}) remains fixed while the total number of vertices $N$ is changed, and therefore the size of holes becomes negligible compared to the surface size in the thermodynamic limit in both cases in  Eq.(\ref{lattices}).  
 
The reason why we assume these two values in Eq.(\ref{lattices}) for $L$ is see that the final results obtained from the model are independent of $L$. In fact, both of the assumed $L$ are negligible to the surface size in the thermodynamic limit, and hence the results should not be independent of $L$ in Eq.(\ref{lattices}).  

The total number of holes in one face of the icosahedron is given by $m(m-1)/2$ and, hence the total number of holes over the surface is $10m(m\!-\!1)$.  Because of the holes on the surface, the total number of vertices $N$ of the lattice with holes are reduced from $N_0$. $(L\!-\!2)(L\!-\!1)/2(\!=\!\sum_i^{L-2} i)$ vertices per a hole are excluded from the surface. Then, we have $N\!=\!N_0\!-\!5(m\!-\!1)(\ell\!-\!m)(\ell\!-\!2m)/m $, because $N$ is given by $N_0\!-\!10m(m\!-\!1) \times(L\!-\!2)(L\!-\!1)/2$, which can also be expressed as $N\!=\!N_0\!-\!5(m\!-\!1)(\ell\!-\!m)(\ell\!-\!2m)/m$ by using $L\!=\!\ell/m$. The lattice size is given by two integers $(\ell,m)$, and hence the sizes $N$ and $L$ are expressed by $(N,L)\!=\!(10\ell^2\!+\!2\!-\!5(m\!-\!1)(\ell\!-\!m)(\ell\!-\!2m)/m ,\ell /m)$. We note that $N$ includes the vertices on the boundary of holes.

The ratio $R_L(\infty)$ of the area of holes to that of the surface including the holes is given $R_L(\infty)\!\to\!(L^2-3)/(2L^2)$ in the limit of $m\!\to\! \infty$ and  $\ell\!\to \!\infty$ under constant $L\!=\!\ell/m$. In fact, the total number of triangles in a hole is $L^2\!-\!3$ and then, the total number of triangles in the holes is $10m(m\!-\!1)(L^2\!-\!3)$, which is easily understood since the total number of faces in the icosahedron is $20$, and the total number of holes in a face is $m(m\!-\!1)/2$ as stated above. On the other hand, the total number of triangles on the triangulated sphere is $2N_0\!-\!4$. Then, we have $R_L(N)\!=\! 10m(m\!-\!1)(L^2\!-\!3)/(2N_0\!-\!4)$. By using $L^2\!=\!(\ell/m)^2$ and $\ell^2\!=\!(N_0\!-\!2)/10$ and identifying the factor $(m\!-\!1)/m$ in $R_L(\infty)$ as $1$, we have $R_L(\infty)\!=\!(L^2-3)/(2L^2)$ in the limit of both $m\!\to\! \infty$ and  $\ell\!\to \!\infty$ with finite $L\!=\!\ell/m$.

\begin{table}[hbt]
\caption{The total number of vertices $N$ and $N_0$ for each $m$, and the corresponding $R_m(N)$. $N_0$ is the total number of vertices of the original lattice, which has no holes.}
\label{table-1}
\begin{center}
 \begin{tabular}{ccccc}
 $(\ell,m)$ & $(N,L)$ & $N_0$   &  $R_L(\infty)$ &  $R_L(N)$  \\
 \hline
  (18,6)   & (2942,3)   & 3242  &   $1/3(\!\simeq\!0.333)$  &   0.278   \\
  (24,8)   & (5202,3)   & 5762  &   $1/3                 $  &   0.292   \\
  (30,10)  & (8102,3)   & 9002  &   $1/3                 $  &    0.3    \\
  (39,13)  & (13652,3)  & 15212 &   $1/3                 $  &   0.308   \\
 \hline
  (20,5)  & (3402,4)   & 4002  &   $13/32(\!\simeq\!0.406)$  &   0.325   \\
  (28,7)  & (6582,4)   & 7842  &   $13/32                 $  &   0.348   \\
  (36,9)  & (10802,4)  & 12962 &   $13/32                 $  &   0.361   \\
  (44,11) & (16062,4)  & 19362 &   $13/32                 $  &   0.369   \\
 \hline
 \end{tabular} 
\end{center}
\end{table}
We show in Table \ref{table-1} some of the numbers that characterize the lattices we use in the MC simulations. In the case of the model in \cite{KOIB-PRE-2007-1}, $m$ is fixed and then the ratio  $R_m(\infty)$ of the area of holes to that of the surface including the holes is given $R_m(\infty)\!=\!(m\!-\!1)/(2m)$ in the limit of $N_0\!\to\! \infty$. In that case in \cite{KOIB-PRE-2007-1} $R_m(N)$ is almost identical with $R_m(\infty)$ even when $N$ is relatively small $N(\sim 10^4)$, while $R_L(N)$ in this paper deviates from  $R_L(\infty)$ about $10\%$ even on the largest surfaces. The reason for the deviation of $R_L(N)$ from  $R_L(\infty)$ is the constraint $m\!\to\! \infty$. In fact, the number $m$ is relatively small compared to $\ell$. The expression $R_L(\infty)$ deviates from $R_L(N)$ just by $(m\!-\!1)/m$. For this reason, the finite size effect influences the model in this paper more strongly rather than the model in \cite{KOIB-PRE-2007-1}.

We note that $R_m(\infty)$ in \cite{KOIB-PRE-2007-1} is $1/3$ and $3/8(\!=\!12/32)$ while $R_L(\infty)$ in this paper is  $1/3$ and $13/32$. Therefore, the two ratios assumed in this paper are both almost identical to those in \cite{KOIB-PRE-2007-1}. The only difference between the two models in this paper and in \cite{KOIB-PRE-2007-1} is in the size of holes.

The canonical Metropolis technique is employed to update the variable $X$ of the model. The variable $X$ is randomly shifted to a new position $X^\prime$ such that $X^\prime\!=\!X\!+\!\delta X$, where $\delta X$ is a position in a small sphere with radius fixed at the beginning of the simulations to maintain about $50\%$ acceptance rate. The new position $X^\prime$ is accepted with the probability ${\rm Min}[1,\exp(-\delta S)]$, where $\delta S\!=\!S({\rm new})\!-\!S({\rm old})$.

\section{Results}\label{Results}
\begin{figure}[htb]
\centering
\resizebox{0.49\textwidth}{!}{%
\includegraphics{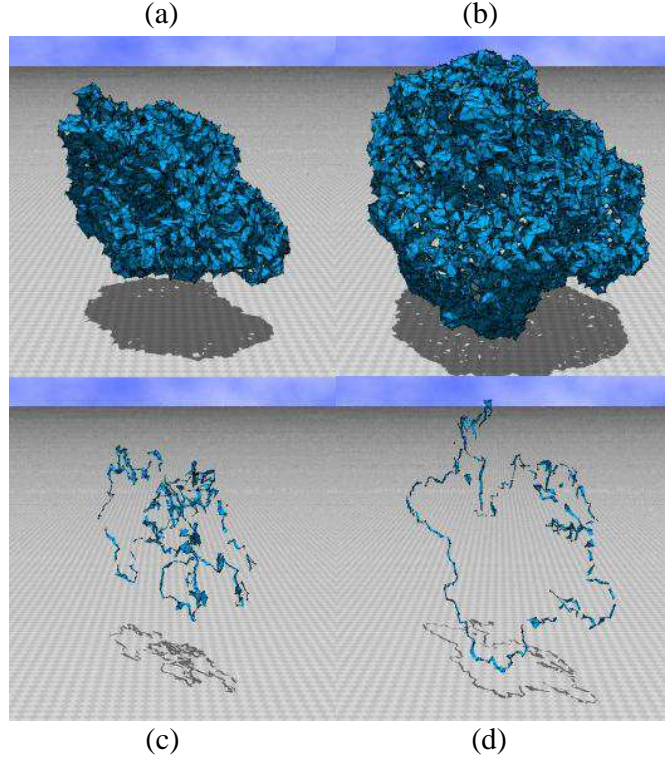}
}
\vspace{0.3cm}
\caption{Snapshots of the surface $(N,L)\!=\!(13652,3)$ in (a) the collapsed phase and (b) the smooth phase at the transition point $b\!=\!1.43$.  (c) The surface section of (a), and (d) the surface section of (b). The mean square size $X^2$ is about $X^2\!=\!63$ in (a), and $X^2\!=\!141$ in (b).  } 
\label{fig-2}
\end{figure}
First we show snapshots of surfaces in Figs. \ref{fig-2}(a) and \ref{fig-2}(b), and the corresponding surface sections in Figs. \ref{fig-2}(c) and \ref{fig-2}(d). These were drawn in the same scale and obtained at $b\!=\!1.43$, which is the transition point of the surface $(N,L)\!=\!(13652,3)$. We should note that two phases, the collapsed and the smooth phases, coexist even on such large surface at $b\!=\!1.43$. The snapshots in Figs. \ref{fig-2}(a) and \ref{fig-2}(b) are typical of the collapsed phase and the smooth phase at the transition point $b\!=\!1.43$. The mean square size $X^2$ of the surfaces in Figs. \ref{fig-2}(a) and \ref{fig-2}(b) is given by $X^2\!=\!63$ and $X^2\!=\!141$, respectively, where $X^2$ is defined as follows:
\begin{equation}
\label{X2}
X^2={1\over N} \sum_i \left(X_i-\bar X\right)^2, \quad \bar X={1\over N} \sum_i X_i,
\end{equation}
where $\bar X$ is the center of mass of the surface.

\begin{figure}[htb]
\centering
\resizebox{0.49\textwidth}{!}{%
\includegraphics{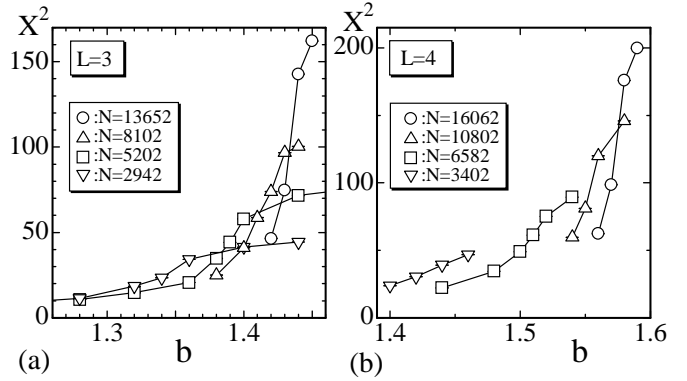}
}
\caption{The mean square size $X^2$ versus $b$ of the surfaces (a) $L\!=\!3$ and  (b) $L\!=\!4$. } 
\label{fig-3}
\end{figure}
The mean square size $X^2$ versus $b$ are shown in Figs. \ref{fig-3}(a) and \ref{fig-3}(b). We see that the transition point $b_c$ moves right with increasing $N$ just like in the model of \cite{KOIB-PRE-2007-1}. The value of $b_c$ is relatively larger than that $b_c\!\simeq\!0.77$ of the model without holes in \cite{KOIB-PRE-2005}, and $b_c$ of the model in this paper is also almost identical to that of the model with many large-sized holes in \cite{KOIB-PRE-2007-1}.

\begin{figure}[htb]
\centering
\resizebox{0.49\textwidth}{!}{%
\includegraphics{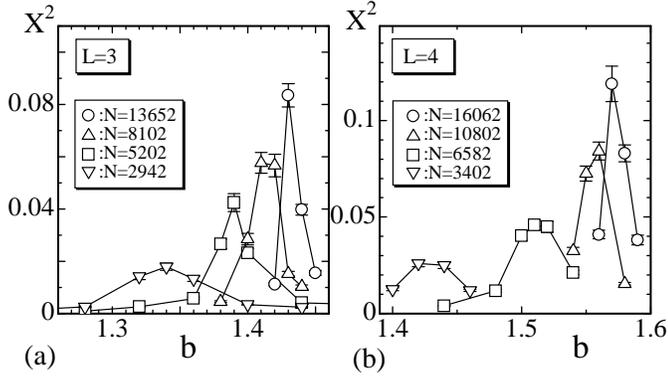}
}
\caption{The variance $C_{X^2}$ versus $b$ of the surfaces (a) $L\!=\!3$ and  (b) $L\!=\!4$.} 
\label{fig-4}
\end{figure}
The variance $C_{X^2}$ of $X^2$ is defined by 
\begin{equation} 
\label{CX2}
C_{X^2} = {1\over N} \langle \; \left( X^2 \!-\! \langle X^2 \rangle\right)^2\rangle,
\end{equation} 
and then size fluctuations are expected to be reflected in $C_{X^2}$. Figures \ref{fig-4}(a) and \ref{fig-4}(b) show  $C_{X^2}$ versus $b$. The anomalous peaks seen in $C_{X^2}$ imply large fluctuations of the surface size and indicate a collapsing transition, just like in the cases of the model without holes \cite{KOIB-PRE-2005} and the model with large-sized holes \cite{KOIB-PRE-2007-1}. The word anomalous denotes the property that the peak value goes to infinite $C_{X^2}^{\rm max}\to \infty$ in the limit of $N\to \infty$. This property will actually be confirmed as a scaling property.  

\begin{figure}[htb]
\centering
\resizebox{0.49\textwidth}{!}{%
\includegraphics{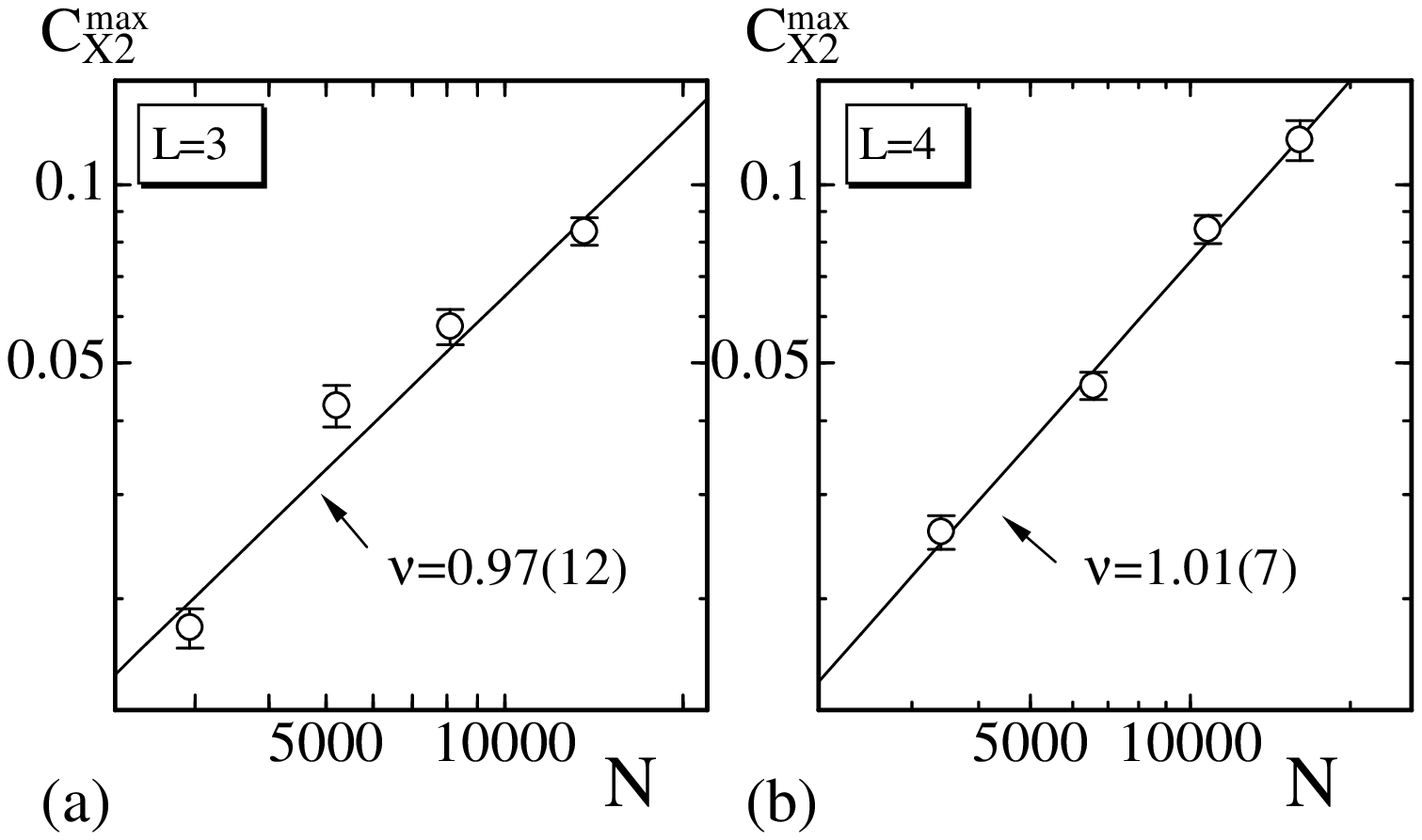}
}
\caption{Log-log plots of the peak value $C_{X^2}^{\rm max}$ against $N$ in the cases (a) $L\!=\!3$ and  (b) $L\!=\!4$. The straight lines were drawn by fitting the data to Eq.(\ref{nu-scale}). } 
\label{fig-5}
\end{figure}
The peak values $C_{X^2}^{\rm max}$ are plotted in a log-log scale against $N$ in Figs. \ref{fig-5}(a) and \ref{fig-5}(b). The scaling property of $C_{X^2}^{\rm max}$ can be seen in the fit of the form 
\begin{equation} 
\label{nu-scale}
C_{X^2}^{\rm max} \sim N^{\nu},
\end{equation} 
where $\nu$ is a critical exponent of the collapsing transition. The straight lines in the figures are obtained by fitting the data to Eq.(\ref{nu-scale}) with the results
\begin{eqnarray} 
\label{nu-value}
\nu_{L=3}=0.97\pm 0.12\quad (L=3), \nonumber \\
\quad\nu_{L=4}=1.01\pm 0.07\quad (L=4). 
\end{eqnarray} 
The finite-size scaling theory indicates that the transition is of first-order in both of the cases $L\!=\!3$ and $L\!=\!4$ because both of the exponents in Eq.(\ref{nu-value}) are $\nu\!\simeq\! 1$ \cite{PRIVMAN-1989-WS}. The fact that $\nu_{L=3}$ and $\nu_{L=4}$ are almost the same within the errors is consistent to the expectation that $\nu$ should be independent of $L$.  

\begin{figure}[hbt]
\centering
\resizebox{0.49\textwidth}{!}{%
\includegraphics{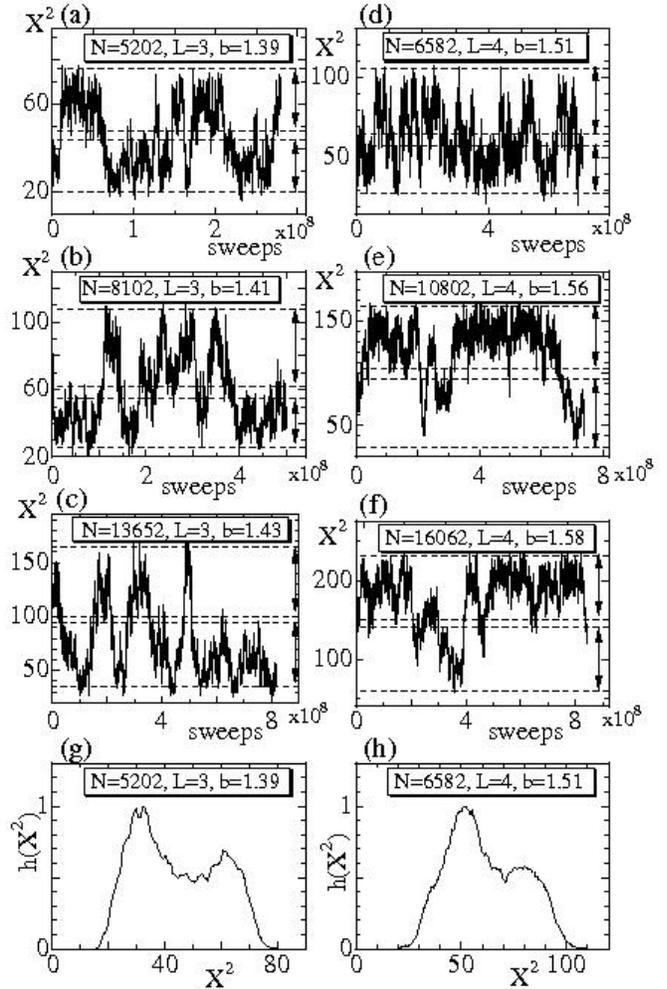}
}
\caption{The variation of $X^2$ against MCS at (a)--(c) $L\!=\!3$, and (d)--(f) $L\!=\!4$. The distribution $h(X^2)$ (histogram) of $X^2$ on the surfaces (g) $L\!=\!3, N\!=\!5202$ and  (g) $L\!=\!4, N\!=\!6582$. The data were obtained at the transition point $b_c(N)$, which depends on both $N$ and $L$. } 
\label{fig-6}
\end{figure}

It is interesting to see whether the collapsed phase is physical in the sense that the collapsed surface can appear in the three-dimensional space. We know that collapsed surfaces of the model with/without holes are characterized by Hausdorff dimension $H<3$ at the collapsing transition point, i.e., the collapsing transition is physical even though the models are just the so-called phantom surface model because of the self-intersecting property. To see the Hausdorff dimension in the collapsed phase at the transition point $b_c$, we firstly show in Figs. \ref{fig-6}(a)--\ref{fig-6}(f) the variation of $X^2$ against MCS (Monte Carlo sweeps) at $b_c$ of the largest three surfaces in both cases $L\!=\!3$ and $L\!=\!4$. Four dashed lines in each figure denote the lower bound $X^{2 \;{\rm col(smo)}}_{\rm min}$ and the upper bound $X^{2 \;{\rm col(smo)}}_{\rm max}$ for computing the mean value $X^2$ in the collapsed phase and in the smooth phase. 

In Figs. \ref{fig-6}(g) and  \ref{fig-6}(h), we plot the distribution $h(X^2)$ (histogram) of $X^2$ obtained at the transition points of the surfaces $L\!=\!3, N\!=\!5202$ and  $L\!=\!4, N\!=\!6582$. Both of the histograms have a double peak structure, which clearly indicates a first-order collapsing transition. The double peak structure can also be expected on larger surfaces such as $N\!=\!13652$ and $N\!=\!16062$, however, it is more hard to see the double peak in the histogram on such large surfaces. 

\begin{table}[hbt]
\caption{ The lower bound $X^{2 \;{\rm col}}_{\rm min}$ and the upper bound $X^{2 \;{\rm col}}_{\rm max}$ in the collapsed state, and the lower bound $X^{2 \;{\rm smo}}_{\rm min}$ and the upper bound $X^{2 \;{\rm smo}}_{\rm max}$ for obtaining the mean value $X^2({\rm smo})$ in the smooth state. }
\label{table-2}
\begin{center}
 \begin{tabular}{ccccccc}
$L$  & $N$ & $b_c$ & $X^{2 \;{\rm col}}_{\rm min}$ & $X^{2 \;{\rm col}}_{\rm max}$ & $X^{2 \;{\rm smo}}_{\rm min}$ & $X^{2 \;{\rm smo}}_{\rm max}$ \\
 \hline
  3  & 2942  & 1.34 & 12 & 24  & 27  & 44  \\
  3  & 5202  & 1.39 & 20 & 44  & 48  & 76  \\
  3  & 8102  & 1.41 & 27 & 55  & 62  & 108  \\
  3  & 13652 & 1.42 & 35 & 93  & 100 & 165  \\
 \hline
  4  & 3402  & 1.44 & 15 & 35  & 39  & 59  \\
  4  & 6582  & 1.51 & 28 & 58  & 65  & 106  \\
  4  & 10802 & 1.56 & 29 & 95  & 105 & 170  \\
  4  & 16062 & 1.58 & 60 & 140 & 150 & 232  \\
 \hline
 \end{tabular} 
\end{center}
\end{table}

 The lower bound $X^{2 \;{\rm col(smo)}}_{\rm min}$ and the upper bound $X^{2 \;{\rm col(smo)}}_{\rm max}$, and some other numbers characterizing the transition are shown in Table \ref{table-2}. The symbol $b_c$ in Table  \ref{table-2} denotes the bending rigidity where the mean values $X^2$ are computed by using the upper and the lower bounds.

\begin{figure}[htb]
\centering
\resizebox{0.49\textwidth}{!}{%
\includegraphics{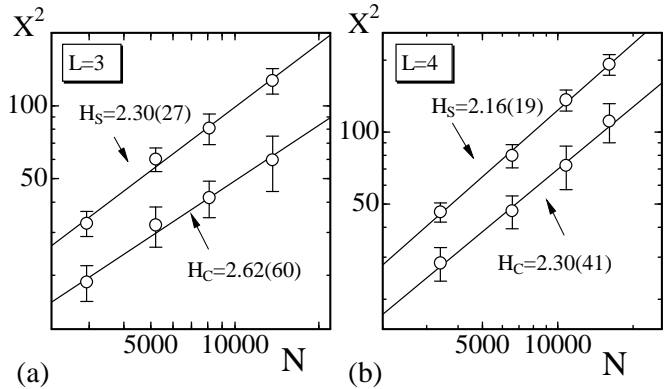}
}
\caption{Log-log plots of the mean values $X^2$ against $N$, where $X^2$ are obtained by averaging the series of $X^2$ with  $X^2_{\rm min}$, $X^2_{\rm max}$ shown in Table \ref{table-2}. } 
\label{fig-7}
\end{figure}
In Figs. \ref{fig-7}(a) and \ref{fig-7}(b) we show the mean values $X^2$ against $N$ in a log-log scale, where $X^2$ are obtained by using the variations of $X^2$ and the upper and the lower bounds at $b_c$ shown in Table \ref{table-2}. The error bars in Figs. \ref{fig-7}(a) and \ref{fig-7}(b) denote the standard deviations. The straight lines drawn in the figure are obtained by fitting the data to the scaling relation
\begin{equation}
\label{Hausdorff-fitting}
X^2 \sim N^{2/H},
\end{equation}
where $H$ is the Hausdorff dimension. We have $H_{\rm smo}$ of the smooth phase and $H_{\rm col}$ of the collapsed phase such that 
\begin{eqnarray} 
\label{H-results}
H_{\rm smo} = 2.30 \pm 0.27,  \; H_{\rm col} = 2.62 \pm 0.60 \quad (L\!=\!3 ),  \nonumber\\
H_{\rm smo} = 2.16 \pm 0.19,  \; H_{\rm col} = 2.30 \pm 0.41 \quad (L\!=\!4 ).
\end{eqnarray} 
We find that $H_{\rm smo}$ in both cases $L\!=\!3$ and $L\!=\!4$ is almost identical to the topological dimension $H\!=\!2$ as expected, and moreover that the collapsed phase is also considered to be physical because  $H_{\rm col}$ is less than $3$ although the errors are relatively large. 

\begin{figure}[htb]
\centering
\resizebox{0.49\textwidth}{!}{%
\includegraphics{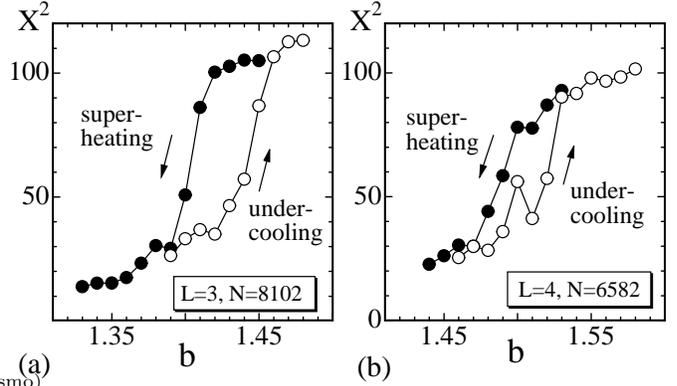}
}
\caption{$X^2$ obtained by undercooling and superheating the surfaces of (a) $L\!=\!3, N\!=\!8102$ and (b) $L\!=\!4, N\!=\!6582$.} 
\label{fig-8}
\end{figure}
Figures \ref{fig-8}(a) and \ref{fig-8}(b) show a hysteresis of $X^2$ obtained in the undercooling and superheating processes on the surfaces of $L\!=\!3, N\!=\!8102$ and   $L\!=\!4, N\!=\!6582$, where the undercooling (superheating) denotes a process with increasing (decreasing) $b$. The starting configuration of the undercooling is a collapsed state, while that of the superheating is a smooth state in each surface. $1\!\times\!10^7$ MCS were done at every value of $b$, and the final configuration obtained at previous $b$ was assumed as the initial configuration of the next $b$ in the processes. The obtained hysteresis is consistent to the first-order collapsing transition, which was confirmed from the double peak structure in $h(X^2)$ in Figs. \ref{fig-6}(g) and \ref{fig-6}(h).

\begin{figure}[htb]
\centering
\resizebox{0.49\textwidth}{!}{%
\includegraphics{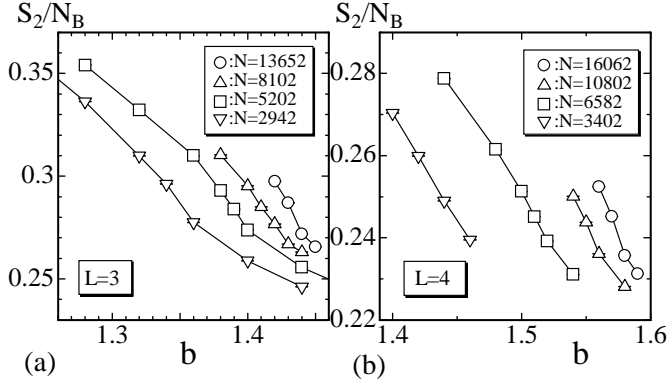}
}
\caption{The bending energy $S_2/N_B$ versus $b$ of the surfaces (a) $L\!=\!3$ and (b) $L\!=\!4$.} 
\label{fig-9}
\end{figure}
Now we turn to the transition of surface fluctuations. The bending energy $S_2/N_B$ is plotted in Figs. \ref{fig-9}(a) and \ref{fig-9}(b) against $b$, where $N_B$ is the total number of bonds excluding the boundary bonds of the holes; $S_2$ is defined only on the internal bonds. The variation of $S_2/N_B$ against $b$ becomes rapid with increasing $N$ and hence is slightly different from that of the model with large-sized holes in \cite{KOIB-PRE-2007-1}, where no transition of surface fluctuations was observed.   

\begin{figure}[hbt]
\centering
\resizebox{0.49\textwidth}{!}{%
\includegraphics{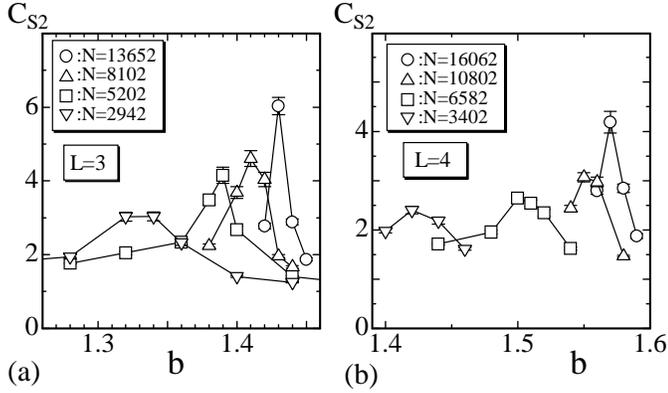}
}
\caption{The specific heat $C_{S_2}$ versus $b$ of the surfaces (a) $L\!=\!3$ and (b) $L\!=\!4$. Anomalous peaks indicate a transition of surface fluctuations. } 
\label{fig-10}
\end{figure}
The specific heat $C_{S_2}$ is defined by the variance of the bending energy $S_2$ such that
\begin{equation} 
\label{CS2}
C_{S_2} = {b^2\over N} \langle \; \left( S_2 \!-\! \langle S_2 \rangle\right)^2\rangle,
\end{equation} 
which is expected to reflect the transition of surface fluctuations. Figures \ref{fig-10}(a) and \ref{fig-10}(b) show $C_{S_2}$ versus $b$. The expected anomalous peaks $C_{S_2}^{\rm max}$ are apparently seen in the figures, and this indicates the transition of surface fluctuations because the height of $C_{S_2}^{\rm max}$ increases with increasing $N$. In the case of the model with small sized holes, we know that $C_{S_2}^{\rm max}$ decreases with increasing $N$ \cite{KOIB-PRE-2007-1}. Therefore, the anomalous structure of $C_{S_2}$ distinguishes the model in this paper and that in \cite{KOIB-PRE-2007-1}, although the size of holes is the only difference between the two models.

\begin{figure}[htb]
\centering
\resizebox{0.49\textwidth}{!}{%
\includegraphics{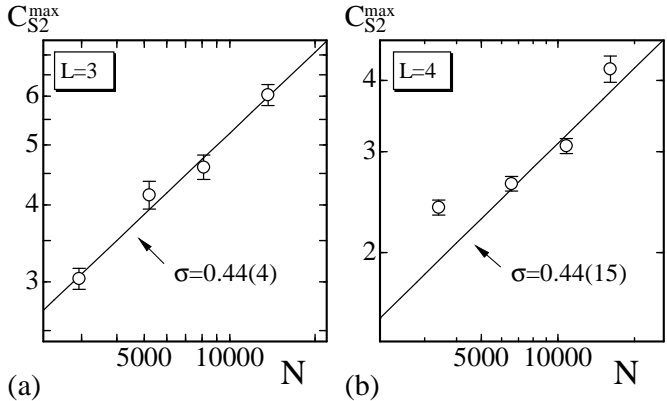}
}
\caption{The log-log plots of $C_{S_2}^{\rm max}$ against $N$ of the surfaces (a) $L\!=\!3$ and (b) $L\!=\!4$. The largest three data in (b) are used in the fitting. } 
\label{fig-11}
\end{figure}
In order to see the dependence of $C_{S_2}^{\rm max}$ on $N$ more convincingly, we plot $C_{S_2}^{\rm max}$ versus $N$ in a log-log scale in Figs. \ref{fig-11}(a) and \ref{fig-11}(b). We see that the data satisfy the scaling relation
\begin{equation} 
\label{sigma-scale}
C_{S_2}^{\rm max} \sim N^\sigma,
\end{equation} 
where $\sigma$ is a critical exponent. The straight lines drawn on the data in Figs. \ref{fig-11}(a) and \ref{fig-11}(b) are obtained by the least squares fitting of the data. The largest three data in Fig.\ref{fig-11}(b) were used in the fitting. The results we obtained are 
\begin{eqnarray} 
\label{sigma-value}
\sigma = 0.44 \pm 0.04\quad (L=3), \nonumber \\
 \sigma = 0.44 \pm 0.15\quad (L=4), 
\end{eqnarray} 
and these indicate a continuous transition in both of the cases $L\!=\!3$ and $L\!=\!4$ because both of the results obviously satisfy $\sigma \!<\! 1$. We should note that the possibility of the first-order transition is not completely eliminated. In fact, the surface size seems insufficient even with $N\!=\!16062$ because of the finite size effects.

\begin{figure}[htb]
\centering
\resizebox{0.49\textwidth}{!}{%
\includegraphics{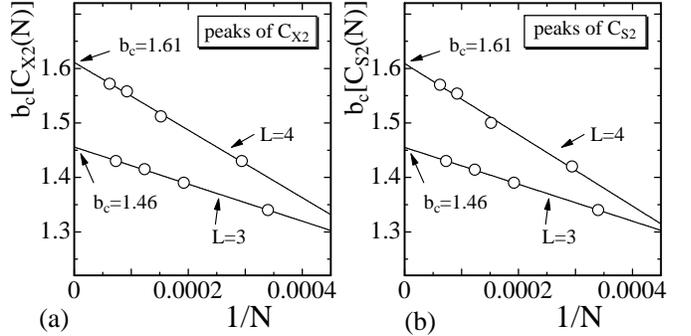}
}
\caption{The transition points $b_c[C_Q(N)]$ vs. $1/N$ for (a) $Q\!=\!X^2$ and (b) $Q\!=\!S_2$. The straight lines are drawn by fitting the data $b_c[C_Q(N)]$ as a linear function of $1/N$. } 
\label{fig-12}
\end{figure}
In the previous section, we mentioned about the finite-size effect in the model of this paper. In order to see this in more detail, we show the transition points $b_c[C_{X^2}(N)]$ and $b_c[C_{S_2}(N)]$ against $1/N$ in Figs. \ref{fig-12}(a) and \ref{fig-12}(b). The symbol $b_c[C_Q(N)]$ denotes the value of $b$ where the variance  $C_Q(N)$ of the quantity $Q$ has its peak. The straight lines are drawn by fitting the data $b_c[C_Q(N)]$ as a linear function of $1/N$. Then we obtain the quantities $b_c[C_Q(\infty)]$ in the limit of $N\to \infty$ such that
\begin{equation} 
\label{bc}
b_c = 1.46 \quad (L=3), \qquad b_c = 1.61 \quad (L=4),
\end{equation} 
where both $b_c$ are common to the quantities $Q\!=\!X^2$ and $Q\!=\!S_2$. 

\begin{figure}[htb]
\centering
\resizebox{0.45\textwidth}{!}{%
\includegraphics{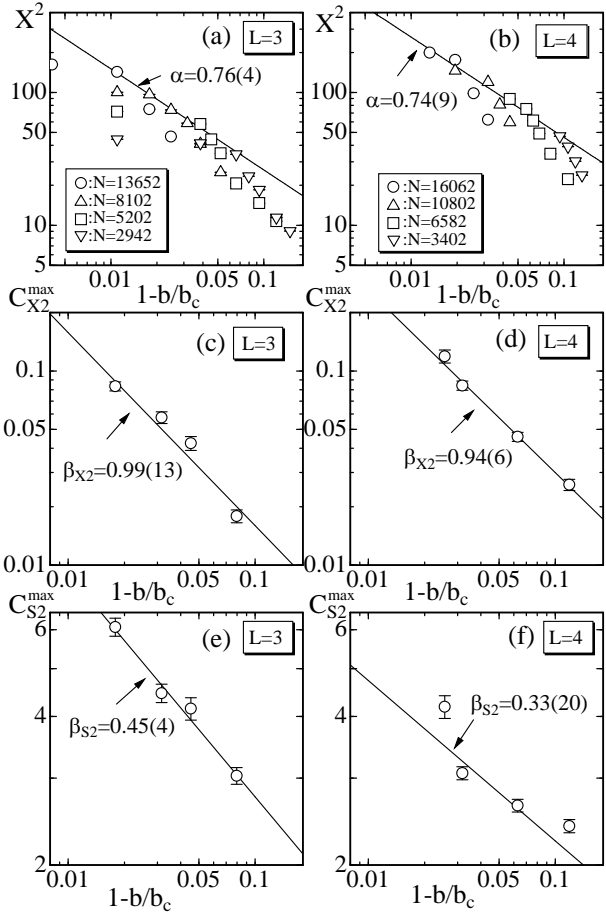}
}
\caption{The log-log plots of $X^2$ against $t=1\!-\!b/b_c$ of the surfaces (a) $L\!=\!3$ and (b) $L\!=\!4$, where $b_c$ denotes the ones in Eq.(\ref{bc}). The straight lines in (a) and (b) are drawn by fitting some of the data obtained in the smooth phase close to the transition point $b_c(N)$ of the surface of size $N$; $X^2\sim t^{-\alpha}$. The log-log plots of $C_{X^2}^{\rm max}$ against $t$ of the surfaces (c) $L\!=\!3$ and (d) $L\!=\!4$, and the log-log plots of $C_{S_2}^{\rm max}$ against $t$ of the surfaces (e) $L\!=\!3$ and (f) $L\!=\!4$. The straight lines in (c)--(e) were drawn by fitting all of the data, and the line in (f) was obtained by using the largest three data; $C_Q^{\rm max}\sim t^{-\beta_Q}$, $Q\!=\!X^2$ or $Q\!=\!S_2$. } 
\label{fig-13}
\end{figure}
Figures \ref{fig-13}(a) and \ref{fig-13}(b) show a scaling property of $X^2$ with respect to $1\!-\!b/b_c$ of the surfaces $L\!=\!3$ and $L\!=\!4$, where $b_c$ is the transition point in the limit of $N\!\to\!\infty$ shown in Eq.(\ref{bc}). Crossover behaviors are almost visible in the figures, however, we concentrate on the smooth phase, which is more clear than the collapsed phase on the figures. The straight lines are obtained by fitting the data $X^2$ in the smooth phase close to the transition point $b_c(N)$ in the surface of size $N$ such that $X^2\sim t^{-\alpha}$, where $t\!=\!1\!-\!b/b_c$. Then, we have the exponents $\alpha$ such that $\alpha_{L=3}=0.76\pm 0.04$ and $\alpha_{L=4}=0.74\pm 0.09$. The two exponents are almost identical within the errors.

The peak values $C_{X^2}^{\rm max}$ and $C_{S_2}^{\rm max}$, shown in Figs. \ref{fig-4} and \ref{fig-10}, can also show the scaling relation $C_{Q}^{\rm max}\sim t^{-\beta_Q}$. We show the log-log plots of $C_{X^2}^{\rm max}$ against $t$ in Figs. \ref{fig-13}(c) and \ref{fig-13}(d) and those of $C_{S_2}^{\rm max}$ against $t$ in Figs. \ref{fig-13}(e) and \ref{fig-13}(f). The exponents $\beta_{X^2}$ of the straight lines in Figs. \ref{fig-13}(c) and \ref{fig-13}(d) can be compared to $\nu$ in Eq. (\ref{nu-value}) and are consistent to the first-order collapsing transition, and  $\beta_{S_2}$ in Figs. \ref{fig-13}(e) and \ref{fig-13}(f) can be compared to $\sigma$ in Eq. (\ref{sigma-value}) and are also consistent to the continuous transition of surface fluctuations. The fitting in  Fig. \ref{fig-13}(f) was done by using the largest three data. Two exponents in Figs. \ref{fig-13}(c) and \ref{fig-13}(d) are identical within the errors, and those in Figs. \ref{fig-13}(e) and \ref{fig-13}(f) are also considered to be identical within the errors. 

Finally, we comment on the value of $S_1/N$, which is expected to be $3(N\!-\!1)/(2N)\!\simeq\!3/2$. Our simulation data satisfy this relation, and hence the simulations are considered to be performed successfully. 

\section{Summary and Conclusion}\label{Conclusion}
In this paper, we have studied the conventional surface model on triangulated spherical lattices with many fine holes, whose size is assumed to be negligible compared to the surface size in the thermodynamic limit. The purpose of the study is to see whether the phase structure is dependent on the size of holes or not in the surface model with many holes.  

Two types of surfaces are investigated: The first is a spherical surface with holes of size characterized by $L\!=\!3$ in the unit of bond length, and the hole size corresponds to $6$ triangles and hence is of hexagonal shape. The second is a surface with holes of size $L\!=\!4$, and the hole size corresponds to $13$ triangles. These holes are the minimum size and the second minimum size in the surfaces constructed in this paper and in \cite{KOIB-PRE-2007-1}. Therefore, the size of holes in the model of this paper is considered to be negligible compared to the surface size in the thermodynamic limit, while the size of holes in \cite{KOIB-PRE-2007-1} is comparable to the surface size in the same limit.

We find that the model in this paper undergoes a discontinuous collapsing transition between the smooth phase and the collapsed phase. Moreover, not only the smooth phase but also the collapsed phase is considered to be physical because the Hausdorff dimensions remain in the physical bound, i.e., $H\!<\!3$, in both phases. This result is identical to that of the homogeneous model in \cite{KOIB-PRE-2005}. These observations are identical to those in the model in \cite{KOIB-PRE-2007-1}, and therefore, we conclude that the collapsing transition is not influenced by the size of holes.

The second observation in this paper is that the model undergoes a continuous transition of surface fluctuations at the same transition point of the collapsing transition. It was reported in \cite{KOIB-PRE-2005} that the homogeneous model undergoes a first-order transition of surface fluctuations. In the case of large-sized holes, no transition of surface fluctuations occurs in the model \cite{KOIB-PRE-2007-1}. Thus, our conclusion is summarized as follows:  The transition of surface fluctuations is influenced by holes in the spherical surface model, and moreover the order of the transition changes depending on the size of holes. 

As mentioned in the Introduction, the free boundary introduced by holes seems to influence the property of the transition of surface fluctuations. Our speculations about the dependence of the order of the transition on the size of holes are as follows: It is possible that the potential barrier between the two degenerate vacuums is dependent on the size of holes, where the vacuum states are both smooth spherical and different from each other only by the orientation. One vacuum state transforms to the other by the symmetry transformation ${\bf n}\to -{\bf n}$. The barrier between the two vacuums becomes lower and lower with increasing size of holes, and the barrier, and consequently the transition, disappears when the size becomes comparable with the surface size \cite{KOIB-PRE-2007-1}. This seems to be connected to the reason why the order of the transition depends on the size of holes.

This work is supported in part by a Grant-in-Aid for Scientific Research from Japan Society for the Promotion of Science.  



\end{document}